\documentclass[prb,aps,showkeys,preprintnumbers,twocolumn,showpacs,amsmath,amssymb]{revtex4}

\usepackage{graphicx}
\usepackage{dcolumn}
\usepackage{bm}
\begin{document}
%
%
\title{Probing spin relaxation in an individual InGaAs quantum dot using a single electron optical spin memory device}
\author{D. Heiss}
\thanks{These authors contributed equally to this work}
\author{V. Jovanov}
\thanks{These authors contributed equally to this work}
\author{F. Klotz}
\author{D. Rudolph}
\author{M. Bichler}
\author{G. Abstreiter}
\author{M. S. Brandt}
\author{J. J. Finley}
\email{finley@wsi.tum.de}
\affiliation{Walter Schottky Institut, Technische Universit\"at M\"unchen, Am Coulombwall 3, D-85748 Garching, Germany}%
\date{\today}
\begin{abstract}
We demonstrate all optical electron spin initialization, storage and readout in a single self-assembled InGaAs quantum dot. Using a single dot charge storage device we monitor the relaxation of a single electron over long timescales exceeding 40$\mu$s. The selective generation of a single electron in the quantum dot is performed by resonant optical excitation and subsequent partial exciton ionization; the hole is removed from the quantum dot whilst the electron remains stored. When subject to a magnetic field applied in Faraday geometry, we show how the spin of the electron can be prepared with a polarization up to 65\% simply by controlling the voltage applied to the gate electrode. After generation, the electron spin is stored in the quantum dot before being read out using an all optical implementation of spin to charge conversion technique, whereby the spin projection of the electron is mapped onto the more robust charge state of the quantum dot. After spin to charge conversion, the charge state of the dot is repeatedly tested by pumping a luminescence recycling transition to obtain strong readout signals. In combination with spin manipulation using fast optical pulses or microwave pulses, this provides an ideal basis for probing spin coherence in single self-assembled quantum dots over long timescales and developing optimal methods for coherent spin control.

\end{abstract}
\maketitle
The spin of charges trapped in semiconductor quantum dots (QDs) is one of the most promising solid state qubits,\cite{1} primarily due to the robust quantum coherence and excellent scaling prospects.\cite{2,3,4,5,6,7,8,9,10,11} Several groups have demonstrated electrical methods to initialize, control and readout QD spin qubits.\cite{2,5,6,12} However, optically active dots are particularly attractive since pulsed lasers can be used to selectively address spin qubits over ultrafast timescales.\cite{9,13,14,15,16} Optical readout of a single spin is extremely challenging and sensitive techniques based on resonant light scattering\cite{17} and Faraday/Kerr\cite{15,18} rotation have been applied. Such measurements are typically repeated at high frequencies ($>$10MHz) to provide enough signal, limiting their potential to probe slow spin dynamics occurring over timescales $>$1$\mu$s. Recent experiments using time-resolved resonance fluorescence showed spin detection\cite{19} in a single QD over much longer timescales. Nevertheless, in such approaches the QDs are tunnel coupled to a Fermi reservoir and co-tunneling leads to enhanced spin relaxation close to the edge of the charging thresholds. Thus, the  electric field regime where the device can be operated is rather limited potentially posing problems when moving from a single dot to few QD systems where electric field also allows tuning of dot-dot tunnel coupling.\cite{20} Here, we demonstrate an all optical spin memory device that allows us to prepare a single spin in an individual QD, store it over millisecond timescales and read it out with high fidelity. Readout is based on spin to charge conversion, whereby spin is first mapped onto the charge status of the dot before charge is repeatedly measured via a luminescence recycling transition. We apply our technique to optically monitor the relaxation of an individual spin in a single dot. Our methods open the way to probe slow spin dynamics, for example due to coupling to nuclear spins or via dipolar interactions with proximal spins.\\
\begin{figure}[htbp]
	\centering
		\includegraphics[width=0.5\textwidth]{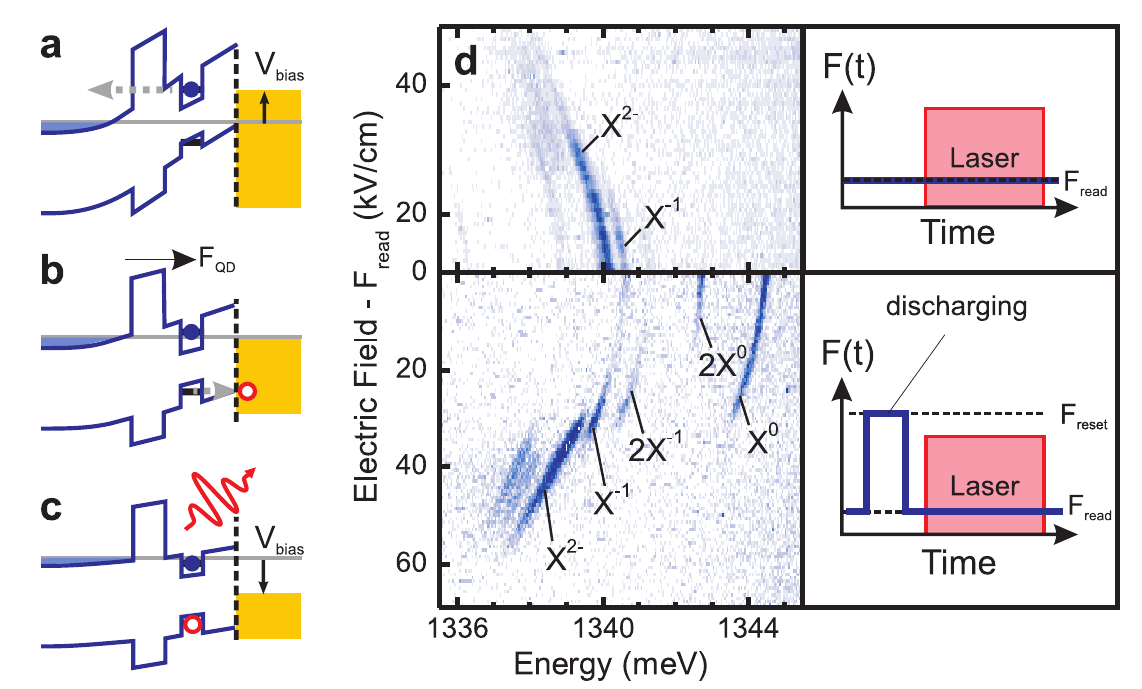}
	\caption{(color online) Schematic representation of the device bandstructure biased in the three operating regimes (a) discharging, (b) optical charging, and (c) readout. (d) Photoluminescence intensity for the neutral exciton ($X^0$), negatively charged exciton ($X^{-1}$, $X^{-2}$) and biexciton ($2X^0$, $2X^{-1}$) emission from a single quantum dot as a function of emitted photon energy and applied electric field with DC bias applied (upper panel) and with application of a periodic discharge pulse (lower panel). The inset shows the voltage sequence.}
	\label{fig:fig1}
\end{figure}
We performed our experiments using a single dot charge storage device (see Refs.~\onlinecite{21,22}). These devices are grown by molecular beam epitaxy and consist of a single layer of self-assembled In$_{0.5}$Ga$_{0.5}$As quantum dots embedded into the intrinsic region of an n-type GaAs Schottky photodiode. After $n^{+}$ buffer layers ($n=1\times10^{18}\rm{cm}^{-3}$) the intrinsic region is formed from the following epitaxial layers in order of growth: 15nm of nominally undoped GaAs followed by a 20nm thick Al$_{0.45}$Ga$_{0.55}$As blocking barrier to facilitate charge storage.\cite{21} This is followed by a 5nm of undoped GaAs onto which a single layer of quantum dots is grown. The epitaxial layer sequence is completed by 100nm of undoped GaAs. An Ohmic contact is established to the buried n-contact  by annealing a GeAuNiAu contact at 420$^{\circ}$C in a forming gas atmosphere. Subsequently, 200x300$\mu$m$^2$ sized windowed photodiodes are fabricated using photolithography and electron beam metalization with a semitransparent 5nm thick Ti-layer. Shadow mask apertures are then defined to optically isolate single dots by evaporating a 30nm thick Ti and 200nm thick Au layer over randomly distributed 1$\mu$m diameter polystyrene balls (Polybeads). These beads are removed after lift-off to produce circular 1$\mu$m diameter apertures in the opaque Au film.

As depicted schematically in Fig. 1, the sample can be biased in three different modes of operation: discharging (Fig. 1a), optical electron charging (Fig. 1b) or readout (Fig. 1c). In the discharging mode the electric field across the QD $F_{QD}$ is tuned such that optically generated electrons and holes are rapidly removed from the dot by tunneling, allowing it to be initially prepared in an uncharged state (Fig. 1a). The lower electric field present in the charging mode (Fig. 1b) ensures that electrons can be selectively created in the dot via optical excitation. Here, the 20nm thick AlGaAs tunneling barrier immediately below the QD layer inhibits electron tunneling escape whilst holes tunnel out rapidly over a timescale determined by $V_{bias}$ the voltage applied. In contrast, during readout (Fig. 1c) $V_{bias}$ is tuned such that the tunneling escape rate of both electrons and holes is much lower than the radiative recombination rate ($\Gamma_{rad}\approx$0.7ns$^{-1}$) and carriers can recombine giving rise to photoluminescence (PL). We identified the discharge, charging and readout modes of operation of our device using optical spectroscopy performed as a function of $V_{bias}$, laser excitation energy and intensity.\cite{21,22} Figure 1d shows $V_{bias}$ dependent PL spectra recorded from a single dot as a function of the electric field applied when the laser is switched on ($F_{QD}$=$F_{read}$). The upper and lower panels of Fig. 1d compare data recorded whilst $F_{read}$ is held constant (upper panel) or with a periodic discharging pulse (lower panel), respectively. Without the discharging pulse a number of negatively charged exciton transitions, labeled $X^{-1}$ and $X^{-2}$ on Fig. 1d, are observed over the entire range of $F_{read}$ investigated. The strong emission from negatively charged excitons and complete absence of $X^0$ emission shows that optical charging takes place for the laser power density and frequency used in the experiment.\cite{21} In contrast, the lower panel of Fig. 1d shows PL recorded from the same dot whilst periodically polling the electric field from $F_{QD}$=$F_{read}$ to $F_{reset}$=200kV/cm for 200ns prior to each optical excitation cycle. In this case, the negatively charged exciton transitions $X^{-1}$ and $X^{-2}$ are accompanied by strong emission from the charge neutral exciton $X^0$ for $F_{read}<$30kV/cm. As shown on Fig. 1d the $X^0$, $X^{-1}$ and $X^{-2}$ transitions from the single dot each have characteristic transition energies  and, thus, the charge state of the QD (0e, 1e or 2e) can be uniquely determined from the PL yield from different excitonic species.

\begin{figure}[htbp]
	\centering
		\includegraphics[width=0.5\textwidth]{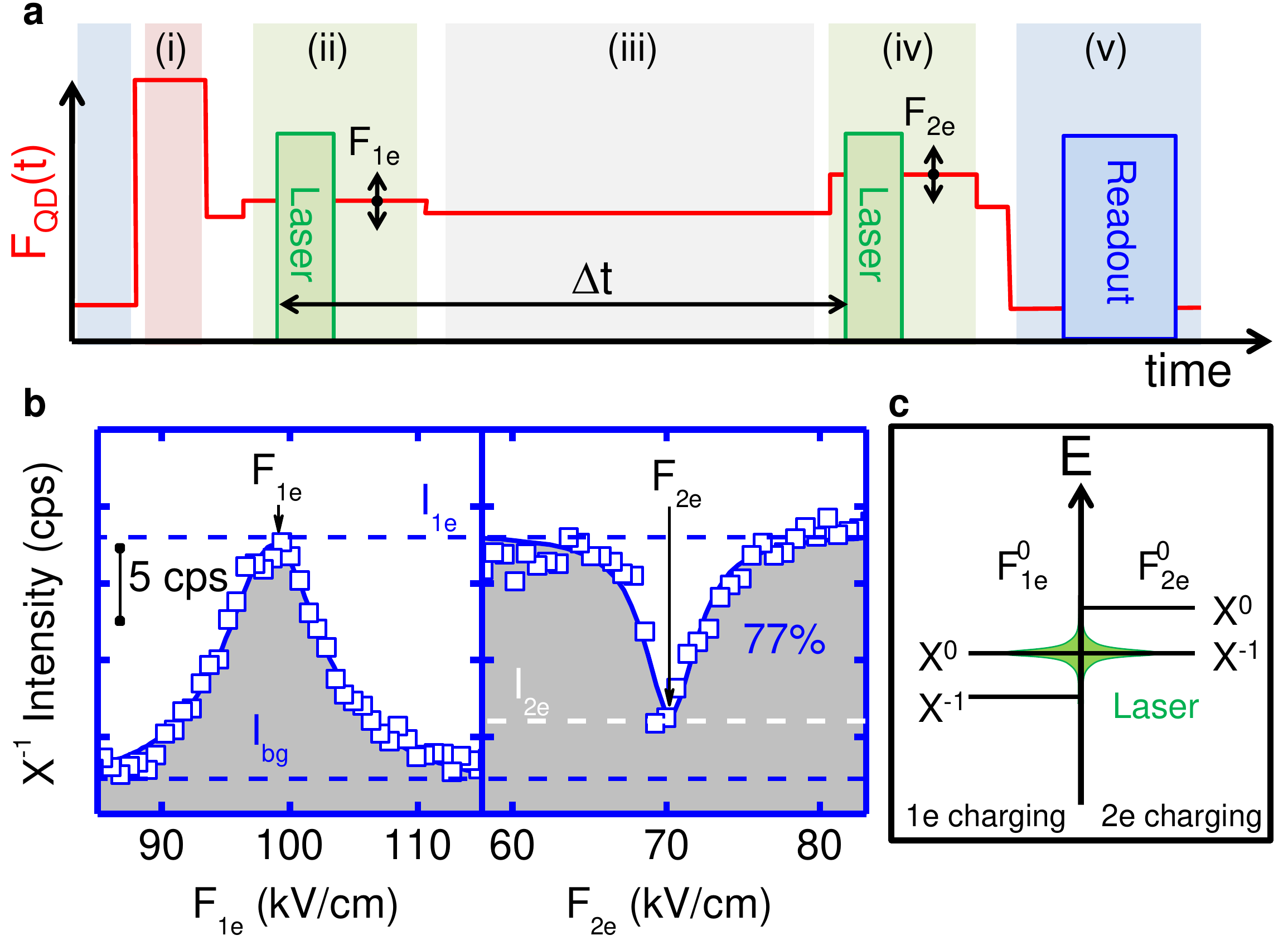}
	\caption{(color online) (a) Electrical and optical pulse sequence used for the single spin measurements. It consists of five distinct measurement phases: (i) discharging, (ii) spin generation, (iii) spin storage, (iv) spin to charge conversion and (v) charge readout. (b) Demonstration of sequential optical charging of the QD with 1e and 2e. The luminescence in the readout phase from $X^{-1}$ is generated as the electric field during the first and the second charging step is varied.}
	\label{fig:fig2}
\end{figure}
Using this information we constructed the electrical and optical pulse sequence used for our single spin measurements. As depicted schematically in Fig. 2a it consists of five distinct phases making up a single measurement cycle: (i) discharging, (ii) spin generation, (iii) spin storage, (iv) spin to charge conversion and (v) charge readout. Initially, in phase (i) of the measurement the dot is emptied of all stored charge.\cite{22} During the spin-generation phase a single electron is optically created in the dot using a single frequency diode laser resonant with $X^{0}$. In practice, the laser frequency is fixed close to the $X^{0}$ resonance and we tune the QD transition into resonance using the DC Stark effect. As depicted schematically on Fig. 2c this resonance occurs at the 1e charging field $F^0_{1e}$. Multiple charging is prevented by the large $X^{0}-X^{-1}$ energy splitting of 4.2meV in our QDs (see Fig. 1d). After retention of the optically generated electron for a storage time $\Delta t$, the spin information is mapped onto the charge occupancy of the dot, either 1e or 2e, by the spin-to-charge conversion phase of the measurement (iv). Hereby, the negatively charged trion transition ($X^{-1}$) is tuned into resonance with the laser at the 2e charging field $F^0_{2e}$. Finally, the sample is biased into the readout mode (Fig. 1c) whereupon the charge occupancy of the dot (1e or 2e) can be probed by measuring the relative luminescence yield from $X^{-1}$. Control measurements show that luminescence can be pumped for at least 100$\mu$s when optically exciting the system via an excited orbital state without changing the charge state of the QD.\cite{21}

To perform our measurements we use a magneto-micro photoluminescence setup that provides a magnetic field up to 12T applied along the QD growth axis and typical sample temperatures of \textit{T}=10K. Resonant optical excitation of the QD is achieved using one of two tunable external cavity single frequency lasers (Sacher GmbH) operating around $\approx$1320meV. The second laser is used to quasi resonantly excite the QD during the readout phase of the measurement via an excited state transition of the QD, approximately ~23meV above the neutral exciton transition. The laser pulses used during the charging phases of the measurement typically had a duration of 300ns and power density of 0.3W/cm$^2$. During the readout phase the second laser is gated on for duration of 1$\mu$s with a power density of 5W/cm$^2$, while simultaneously turning on a single photon counter (Perkin Elmer SPCM-AQR-16). The control voltage sequence was applied to the sample by an arbitrary waveform generator (Agilent 33250A) that is synchronized to the laser charging 1e, 2e and readout pulses (see Fig. 2a). The switching speed of the device is limited to 100ns by stray capacities and impedance mismatch in the experimental setup. Nevertheless, the timing of the voltage sequence and the laser pulses has been adapted to allow voltage stabilization before each of the laser pulses are turned on. The data presented in this manuscript is typically integrated over 10$^6$ measurement cycles.

In first experiments we begin by applying the pulse sequence depicted in Fig. 2a whilst scanning only the electric field applied during the 1e charging phase of the measurement ($F_{1e}$) and blocking the laser during the 2e charging phase. As $F_{1e}$ is swept, we monitor the integrated intensity of the singly charged states ($X^{-1}$ and $2X^{-1}$) during the readout phase of the measurement. As $F_{1e}$ is tuned through $F^0_{1e}$=99kV/cm we observe a clear peak in the $X^{-1}$ emission intensity (Fig. 2b, left panel), whilst keeping the readout phase of the measurement completely unchanged, showing that the dot is successfully charged with a single electron. We then test whether a second charge can be subsequently added. To do this, we unblock the laser during the 2e charging phase of the measurement (iv) and scan $F_{2e}$ to achieve resonance between the excitation laser and $X^{-1}$. Figure 2b shows a measurement where the pulse sequence displayed in Fig. 2a is used including both the 1e and 2e charging steps. The first charging field $F^0_{1e}$=99kV/cm is fixed to charge with a single electron, while $F_{2e}$ is scanned from 50-85kV/cm to search for the resonance between $X^{-1}$ and the charging laser. A clear dip is observed close to $F^0_{2e}$=70kV/cm with a reduced $X^{-1}$ PL signal showing that a second electron has been added to the dot (Fig. 2b, right panel). The relative intensity $r=(I_{1e}-I_{2e})/(I_{1e}-I_{bg})$=0.77 ($I_{1e}$ and $I_{2e}$ being the $X^{-1}$ intensities at 1e and 2e charging fields and $I_{bg}$ being the background signal) reflects the efficiency with which we add a second electron to the dot.

Until now the spin of the first and second electrons were not considered. At high magnetic fields and linear polarized excitation, the Zeeman splitting of the $X^{0}$ line can be spectrally resolved ($|g_{ex}|$=0.6) providing a convenient way to select the spin of the first electron generated in the dot and probe its spin projection after a storage time $\Delta t$. Each Zeeman branch can be selectively excited simply by fixing the energy of the charging laser and identifying the charging fields for spin up ($F_{1e,\uparrow}$) and spin down ($F_{1e,\downarrow}$) electrons, respectively. A typical measurement performed at a magnetic field $B$=12T is presented in Fig. 3a, clearly showing two charging peaks corresponding to the two different stored electron spin states. The spectrum can be fitted with two Lorentzian peaks, having an energy linewidth of 0.4meV, centered at $F_{1e,\uparrow}$=87kV/cm and $F_{1e,\downarrow}$=91kV/cm, respectively. The identification of the spin states is verified by the spin dynamics presented below.

\begin{figure}[htbp]
	\centering
		\includegraphics[width=0.5\textwidth]{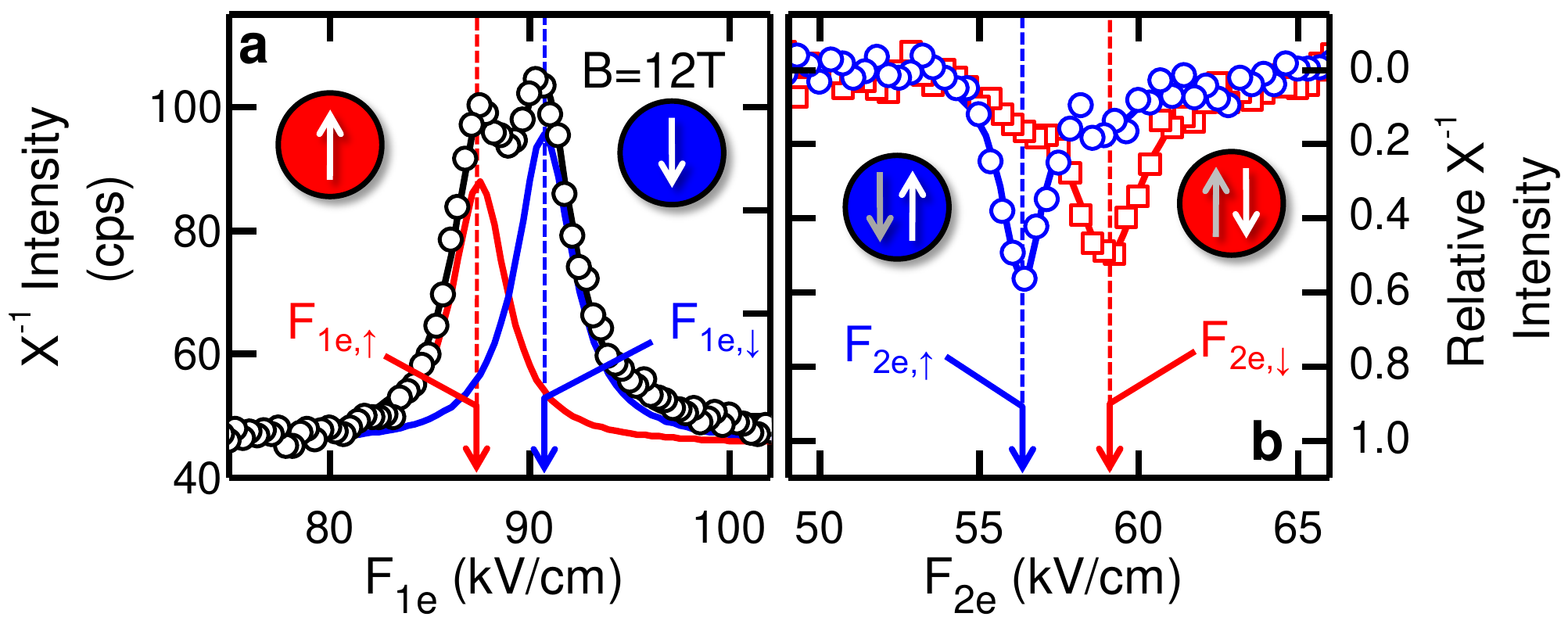}
	\caption{(color online) (a) Measurement of the 1e charging characteristics with a magnetic field $B$=12T applied in growth direction, for which the Zeeman components of the neutral exciton can be clearly resolved. The Zeeman split absorption lines can be attributed to the creation of spin up (down) electrons at electric fields of $F_{1e,\uparrow}$ and $F_{2e,\uparrow}$ ($F_{1e,\downarrow}$ and $F_{2e,\downarrow}$). (b) A clear suppression of the charging efficiency is detected for addition of parallel spins in the 2e charging step, demonstrating spin to charge conversion.}
	\label{fig:fig3}
\end{figure}
The selective generation of a spin orientated electron in the QD is confirmed by measuring the spectral characteristics of the 2e charging dip (see Fig. 3b) and its evolution with the storage time ($\Delta t$) between the 1e and 2e charging phases of the measurement (Fig. 2a). Figure 3b shows the 2e charging dips measured following excitation of a spin up ($F_{1e,\uparrow}$=87kV/cm - squares) or spin down ($F_{1e,\downarrow}$=91kV/cm - circles) electron in the dot, respectively. Upon exciting the first electron spin up and storing it for 800ns, we observe a pronounced 2e charging dip at $F_{2e,\downarrow}$=58kV/cm (squares - Fig. 3b). In strong contrast, upon exciting the first electron spin down we observe the 2e charging dip at $F_{2e,\uparrow}$=56.5kV/cm (circles - Fig. 3b). Mapped onto an energy scale, the splitting between the $F_{2e,\uparrow}$ and $F_{2e,\downarrow}$ 2e charging dips correspond precisely to the observed Zeeman splitting of the 1e charging peaks. This is expected since the Zeeman splitting of $X^{0}$ and $X^{-1}$ are identical.\cite{23} These observations clearly show that the spin projection of the first electron generated in the dot can be directly measured via the relative amplitudes of the 2e charging dips. Moreover, they demonstrate that the spin projection is retained across the storage phase of the measurement, before the 2e charging step is attempted.

To gain more insight on the selectivity of the spin generation, we performed measurements of the initial degree of spin polarization $\rho=r_\uparrow - r_\downarrow$ as a function of $F_{1e}$. The degree of polarization is presented in Fig. 4. The measurement was performed using a storage time of $\Delta t$=0.8$\mu$s. To compensate for relaxation effects the degree of polarization has been extrapolated to $\Delta t$=0 using the relaxation dynamics detected in time resolved experiments presented later. The minimum and maximum of $\rho$ are found around $F_{1e,\uparrow}$=87kV/cm and $F_{1e,\downarrow}$=91kV/cm, with values of 60\% and 65\%, respectively. Between these two maxima a sharp transition is observed with a change of sign, whilst towards lower and higher electric fields, the degree of polarization decreases to zero. With increasing detuning from the charging resonance, the 1e charging efficiency reduces rapidly and, thus, the fidelity of the measurement degrades. The efficiency of spin up and spin down generation can be estimated by fitting the 1e charging peak with two Lorentzian peaks (Fig. 3a). The maximum degree of polarization for undisturbed spin generation is then calculated from these efficiencies and shown as full line in Fig. 4. Since the calculated values coincide well with the measured degree of polarization, we conclude that pure spin initialization is only limited by the selective addressing of a single Zeeman level. Spin initialization can, therefore, either be improved by using quantum dots with larger Zeeman splitting or by the use of circular polarized light. These observations clearly demonstrate that the device operates as an optically addressable single electron spin memory.

\begin{figure}[htbp]
	\centering
		\includegraphics[width=0.5\textwidth]{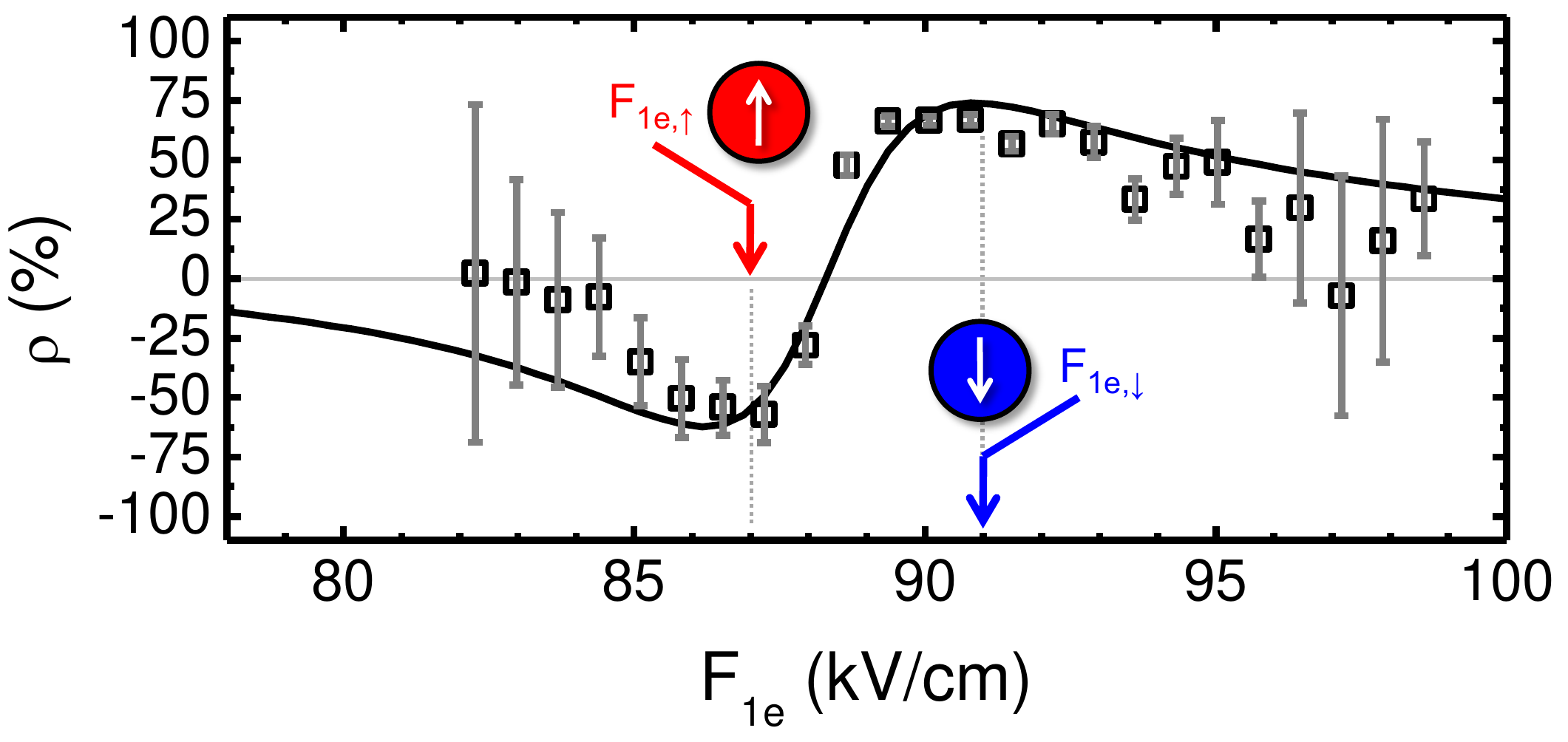}
	\caption{(color online) The degree of polarization $\rho=r_\uparrow - r_\downarrow$ measured as a function of the first charging field $F_{1e}$. The maximum values of $\rho$ are limited by the spectral separation of the two excitonic Zeeman levels.  The full line shows the expectation based on two Lorentzian transitions with a spectral linewidth of 0.4meV, determined by the tunneling time of the hole out of the dot. The good agreement indicates that the fidelity of the spin preparation is limited only by the relatively low excitonic g-factor in this sample.}
	\label{fig:fig4}
\end{figure}
We now apply our methods to monitor the spin relaxation of a single, optically prepared electron. Again, the pulse sequence depicted in Fig. 2a is used and $\Delta t$ is varied from 0.8$\mu$s to 16$\mu$s by changing the delay between the 1e and 2e charging pulses. The resulting data is presented in Figs. 5a and 5b.  The 1e charging electric field $F_{1e}$ is chosen such as to initialize spin down (Fig. 5a) or spin up (Fig. 5b) electrons, respectively. The data is presented as probabilities $p_{\uparrow,\downarrow} = r_{\uparrow,\downarrow}/(r_{\downarrow} + r_{\uparrow})$ to find the electron in a particular spin orientation after the storage time $\Delta t$. For short storage times, a dominance of the spin orientation that was chosen for initialization is observed. As the storage time increases the system evolves towards thermal equilibrium, which can be described by the Boltzmann statistics of non-interacting spins. In all four traces of Figs. 5a and 5b a mono-exponential decay can be fitted with a single time constant, the spin relaxation time $T_1$=3.1$\pm$0.4$\mu$s. This value is in excellent agreement with lifetime measurements performed under similar conditions on ensembles of self-assembled QDs,\cite{11} showing that relaxation is mediated by spin-orbit coupling mediated by single phonon scattering.\cite{4,10,11}

\begin{figure}[htbp]
	\centering
		\includegraphics[width=0.5\textwidth]{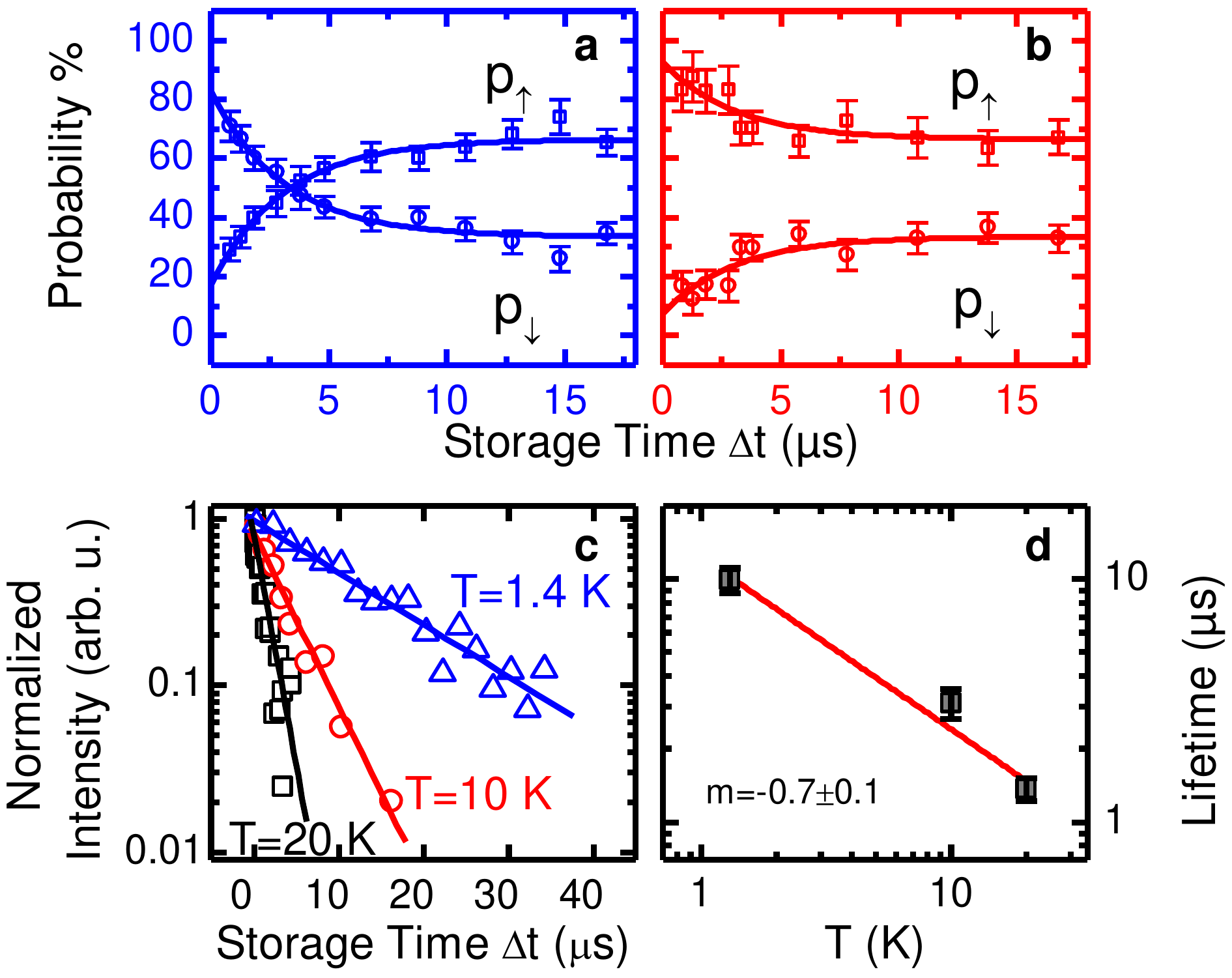}
	\caption{(color online) (a, b) Time evolution of the probabilities for a particular electron spin orientation after spin down (a) and spin up (b) initialization. An electron spin relaxation time of $T_1$=3.1$\pm$0.4$\mu$s is extracted from this data. (c) Temperature dependence of the 2e charging efficiency after initialization of a spin up electron. The data is presented as ($r_\uparrow-r_\downarrow$) normalized to the initial value on a semi-logarithmic scale. (d) Extracted spin relaxation times fitted to the data presented in (a). As expected, relaxation times decrease with increasing temperature.}
	\label{fig:fig5}
\end{figure}
Similar measurements performed as a function of lattice temperature reveal that $T_1$ decreases with increasing temperature, from 9.9$\pm$1.1$\mu$s at $T$=1.4K to 1.4$\pm$0.2$\mu$s at $T$=20K (Fig. 5c). Note that in the case of the low temperature measurements an extremely long storage time of 34$\mu$s is used, during which reliable spin detection is still possible. A power $T_1 \propto T^m$ law with $m$=-0.7$\pm$0.1 is fitted (Fig. 5d), in agreement with the expected linear behavior for phonon mediated spin-orbit coupling of the two spin states in the dot.\cite{5}

In summary, we have demonstrated a single QD optical spin memory device and used it to monitor spin relaxation of an individual electron over timescales approaching a millisecond. We showed that the spin of the optically prepared electron can be selected simply by choosing the gate voltage applied to the device and, moreover, read out the spin state using an all optical implementation of a spin to charge conversion technique. When compared with the spin storage technique employed previously,\cite{11} where each spin generated one photon per measurement cycle, our results represent an increase of four orders of magnitude in photons per spin, per measurement cycle. This was made possible by mapping the spin information onto the more robust charge status of the QD and then measuring charge. Whilst our methods already allow us to probe an individual spin in a single QD, with further optimization of the optical collection efficiency and sample design, single shot spin readout is conceivable. Finally, we note that perturbations arising from illumination, DC currents and co-tunneling phenomena are absent in our experiment whilst the spin is stored in the QD. When used in combination with fast optical spin manipulation or microwave pulses, this builds an ideal basis for probing spin decoherence in single self-assembled QD over long timescales and developing optimal methods for coherent spin control.

The authors gratefully acknowledge financial support by the DFG via SFB631, the German Excellence Initiative via the Nanosystems Initiative Munich (NIM) and the European Union via SOLID.

\end{document}